\documentclass[11pt]{article}
\usepackage{graphicx}
\begin{document}
\title{Critical or Tricritical Point in 
Mixed-Action SU(2) Lattice Gauge Theory?}
\author{Michael Grady\\
Department of Physics\\ State University of New York at Fredonia\\
Fredonia NY 14063 USA}
\date{\today}
\maketitle
\thispagestyle{empty}
\begin{abstract}

An analysis of scaling along the first-order bulk transition line
in fundamental-adjoint SU(2) lattice gauge theory 
strongly supports the first-order endpoint being a tricritical point, 
and is inconsistent with it being an ordinary critical point
as is usually assumed. 
If tricritical, the transition must continue from the endpoint
further into the phase diagram as a second-order bulk transition and extend
to and beyond the Wilson axis. Observations 
indicate that this is most likely the same transition that
has been traditionally considered a finite-temperature transition. 
\end{abstract}
\section{Introduction}
\begin{figure}[ht]
                       \includegraphics[width=5in]{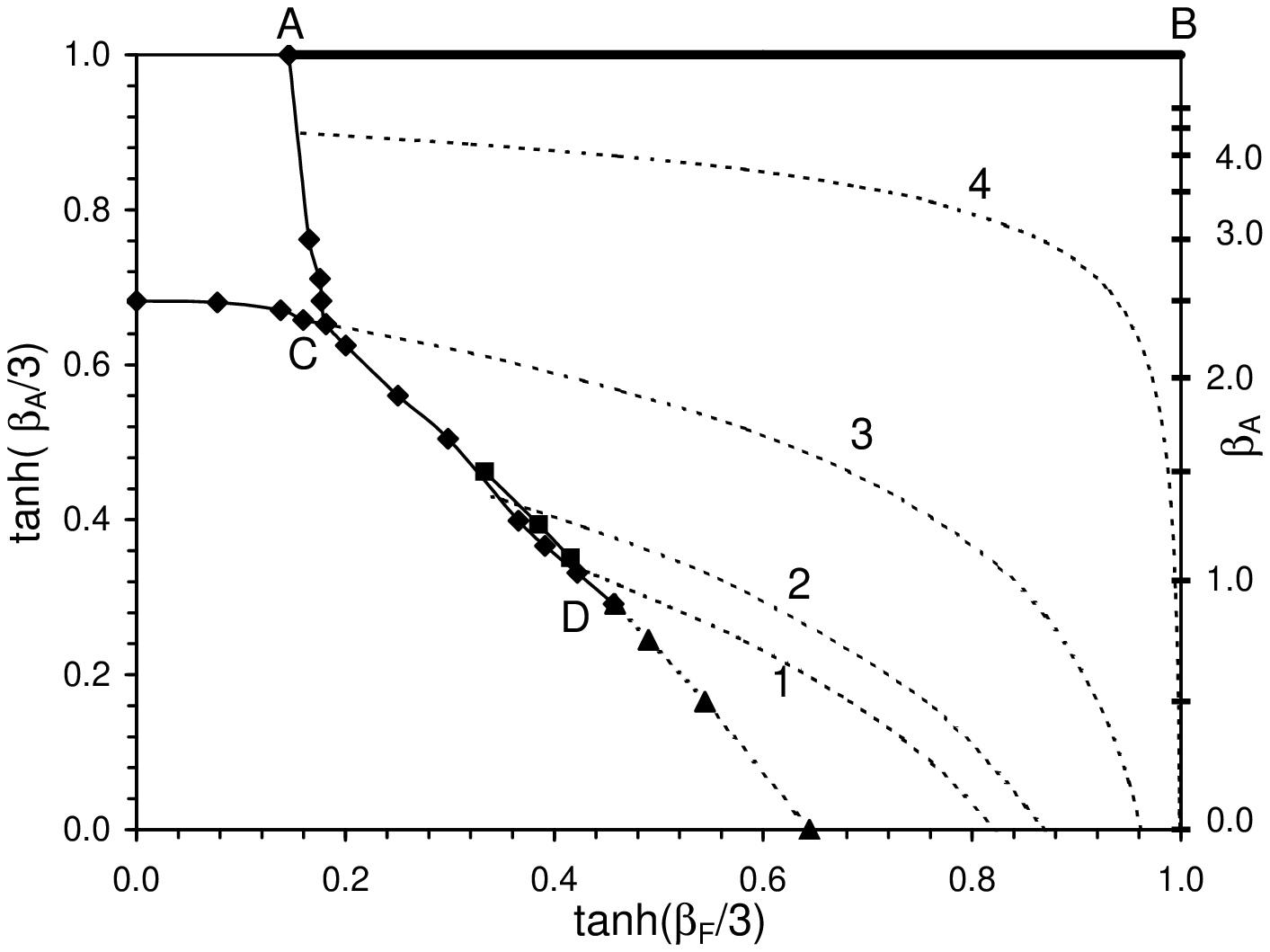}
            \caption{Phase transitions on the fundamental-adjoint plane. 
Nonlinear
axes have been chosen
so the entire coupling plane, including continuum limit at right and Z2 
lattice gauge theory at top can be seen. 
Scaling of couplings has been chosen to
reproduce ``look" of usual plot on linear axes. 
Nonlinear scale for $\beta _A$
is shown at right. Diamonds are Bhanot-Creutz data \cite{bc}, 
triangles are second-order $8^3 \times 4$, and squares are
$8^3 \times 4$ first-order data from \cite{ggm}. Solid lines are
first-order, dashed second-order. Lines 1-4 are hypothetical 
second-order lines for very large lattices, following perturbative
lines of constant physics. The line $\overline{\rm{AB}}$ is the deconfined 
phase of the 4-d Z2 lattice gauge theory. }
          \label{fig1}
       \end{figure}

The characterization of phase transitions has often been made clearer by
considering  higher-dimensional coupling spaces, especially ones that 
become more-familiar or exactly-known theories 
at one or more edges of the phase diagram. Then 
one can see how the various 
phase boundaries and critical points attach to better-known transitions.
In SU(2) and SU(3) lattice
gauge theory, the fundamental-adjoint plane has provided interesting
insights. The SU(2) case was first studied by Bhanot and Creutz \cite{bc},
who found two lines 
of first-order transitions which joined at a triple point 
and then continued as 
a single first-order line until ending at a 
presumed critical point (Fig.~\ref{fig1}). They argued that
since the transition apparently ended, 
the strong coupling confining phase
could be continued around the endpoint resulting in a 
confining continuum limit,
since a connecting path which encounters no phase 
transition could be found. 
The fact that the Polyakov loop, an order parameter for deconfinement 
(or disorder parameter for confinement), appeared to undergo a sudden 
jump to non-zero values across this line would seem to be inconsistent
with the idea that both sides of the transition were confining, 
however it was assumed that
in the limit of an infinite lattice the 
deconfinement signal would disappear. 
One place where it cannot disappear, 
however is along the top line of the phase diagram
where $\beta _A = \infty$. 
This is the well-know Z2 lattice gauge theory, which has a
bulk first-order transition at
$\beta _F=\frac{1}{2} \ln (1+\sqrt{2}) \approx 0.44$ 
(determined exactly from self-duality)\cite{wegner}. 
This transition is deconfining with the Polyakov loop 
as order parameter. Therefore the line $\overline{\rm{AB}}$ on
Fig.~\ref{fig1} is definitely deconfined even on the 4-d infinite lattice. 
In the conventional interpretation of lattice gauge theory 
the entire rest of the phase diagram
is confining on such a lattice.

The situation became clearer when it was realized that finite
lattices were at a finite
3-d (ordinary physical) temperature which increased 
as $\beta$ increased. A deconfinement 
transition due to physical temperature - a so called 
finite-temperature transition - should 
occur on a finite lattice. For SU(2) this is a 
second-order transition and was studied 
extensively on the Wilson axis ($\beta _A =0$). 
In Ref.~\cite{ggm}  
the finite-temperature transition
was studied on the fundamental-adjoint plane on lattices 
with temporal extent $N_\tau = 4$.
Some couplings were further studied at $N_\tau = 6$ and 8 \cite{rajivt6}.
The rather surprising result of these studies was 
that the line of second-order finite-temperature transitions seemed to 
join up with the first-order bulk transition
at its endpoint. 
The finite-temperature deconfinement
transition also became first-order at this point 
(point D in Fig.~\ref{fig1}). If these 
transitions truly were joined that would call 
into question the finite-temperature interpretation of
the second-order transition. This is because a finite-temperature 
transition should move all
the way to the right of the phase diagram as $N_\tau$ is 
increased (similar to the hypothetical 
dashed lines 1-4
on Fig.~\ref{fig1}). However this would not be possible if 
one end was tied down at point D. In this case it 
would be unlikely for the second-order transition to move 
beyond line 1 (this is a line of constant
physics as determined by continuum two-loop perturbative
renormalization-group theory, as are lines
2-4). Line 1 is constructed to join at the endpoint
of the first-order line,
as determined by extrapolating the latent heat (which appears to
vary linearly with $\beta _A$)
to zero using the $12^4$ lattice
data from the current study. 
This is at ($\beta _F$, $\beta _A$)=(1.38$\pm 0.03$, 1.04$\pm 0.01$), 
a somewhat
higher $\beta _A$ (and lower $\beta _F$) than the original 
low-statistics Bhanot-Creutz result. This is in agreement
with the results of Refs. \cite{ggm,rajivt6}, 
although not with \cite{rajivsl} (more on this later). In the
current paper,
this point will be referred to as the ``first-order-endpoint" (FOE), as its
identification as either a critical point or a 
tricritical point is the question
under consideration. 
To the right of the phase line the coupling quickly becomes weak enough
for these perturbative lines of constant physics to be accurate.  
Near the phase transition they are more hypothetical. 
Putting aside the unreasonable possibility of a phase line
that curves up and then down, one can therefore 
conclude that if the second-order
deconfinement transition continues to join 
the bulk transition at point D as 
$N_\tau \rightarrow \infty$ then,
at least for $\beta _A >1$, the zero-temperature continuum limit
(right hand side of the phase diagram) would be deconfined. 
In other words, the entire transition would
be bulk. The movement of the second-order line 
with changing lattice size, seen on the Wilson axis,
would be explained as an ordinary finite-size shift in the
critical point, perhaps with an unusually large shift-exponent or 
following something other than a power law. 
In this case the transition point would converge to some finite
value of $\beta _F$, probably in the range 3.0 to 4.0, 
as $N_\tau \rightarrow \infty$.

  Another hypothesis consistent with the 
original Bhanot-Creutz scenario is that the two transitions
join, but {\em not} at the FOE. 
There are a few systems known in which a 
second-order line meets a first-order line
at a point other than its endpoint, such as the 
Blume-Emery-Griffiths model\cite{beg}
and certain metamagnets\cite[pp175-181]{CL}. 
These systems each 
have two different order parameters
and two corresponding correlation lengths. However, in these systems both
transitions are bulk.

Applying this scenario to the lattice gauge theory, 
where the second-order transition
is finite-temperature,
the point where the second-order transition joins the 
first-order line is hypothesized to slowly move up the diagram as $N_\tau$ 
increases (lines 2-4 in Fig.~\ref{fig1}). The 
lower part of the bulk transition would no longer be deconfining. 
Eventually as $N_\tau$ became infinite
the entire phase diagram except for the line AB would be confined. 
The bulk transition would have 
nothing to do with confinement. The observed deconfinement across
the bulk line on small lattices would be due to
the coincident finite-temperature transition, 
which somehow becomes first-order 
due to the influence of the bulk transition.
The bulk transition  would have its own order 
parameter which was not symmetry-breaking,
similar to the liquid-gas transition. 
This order parameter would have a correlation length
associated with it which would become infinite at the critical point 
at the end of the first-order line.
This would happen at a place within the confining region, 
where the string tension 
(and correlation length associated with it) is finite. In other 
words the theory would have to have two independent correlation lengths. 
This seems a bit odd in that there
is no evidence for the more-complicated 
scaling laws that would normally result from a theory
with more than one correlation length. 
However, otherwise this interpretation is consistent with the 
conventional interpretation of Lattice Gauge Theory - 
in particular with a confining 
zero-temperature continuum limit.

  It is difficult to distinguish these two 
hypotheses simply by looking for a joining away from the 
FOE, because one would have to go to rather high $N_\tau$ to get a 
convincing separation. A small 
separation was reported for the SU(3) case \cite{su3} which 
has a similar phase diagram, except that both transitions
(bulk and deconfining) are first-order. 
However, critical points determined on finite lattices from different
quantities or by different techniques 
can differ substantially from one another. 
This could lead to a small apparent separation of
the critical point and the end of the finite-temperature line, 
since one is determined from the plaquette and 
the other from the Polyakov loop. A more convincing demonstration 
of separation would be the observation of uncorrelated tunneling events
in two different order parameters. 

  There is actually a much easier way to determine which 
of these cases is correct, based on a study of the 
bulk transition itself. In the conventional scenario just described, 
the line of first-order transitions
ends in an ordinary critical point. 
Its order parameter therefore cannot be associated with spontaneous
symmetry breaking, because otherwise the 
transition would have to continue, in order 
to divide the plane into symmetry-broken and unbroken parts. 
First-order transitions ending in a critical point
are characterized in the Landau theory as having a 
cubic term in the free energy which explicitly breaks
the symmetry. Only at the critical point, where the 
cubic term is zero, is the symmetry accidentally realized,
allowing for a single point of criticality. 
The other scenario, where the second-order line joins the 
first-order at the endpoint, is exactly what happens at 
a {\em tricritical} point. A tricritical point is 
associated with a symmetry-breaking phase transition. 
Its Landau free-energy has only even-order terms
but must be considered out to sixth order, 
because in part of the phase diagram the quartic term is negative.
Here there is a first-order transition, 
which becomes second-order when the quartic term becomes positive.
The tricritical point is simply where the change in order takes place, 
when the quartic term vanishes.
If the endpoint of the fundamental-adjoint
bulk transition is tricritical, then the transition {\em must} 
continue as second-order, 
and it must be symmetry breaking. 

  A favorable aspect of this study is that the 
bulk transition should not, by its very nature, depend much
on lattice size. Bhanot and Creutz's study was done on a 
$5^4$ lattice, and more recent results, 
such as those presented here
for $12^4$ and $20^4$ lattices, do not differ much 
in the location of the transition
or other parameters such as latent heat. 
Except for an expected reduction in variance from simple statistics, no 
significant differences are seen between our runs 
on $12^4$ and $20^4$ lattices(see Fig.~\ref{fig3} below).
Similarly, because it is a bulk quantity,
one would expect the Landau free energy function 
to be accurately determined by modest lattices with
very minor corrections from surface effects. 
Any results linked to the behavior of this free energy
are therefore unlikely to change much on larger lattices. 

The lack of finite size dependence seen in the data shown 
below contrast with what was reported by Gavai who also studied the
bulk transition on symmetric lattices\cite{rajivsl}. This 
can be traced to a difference in measurement
technique. Gavai found a significant decrease in latent heat as lattice
size was increased at $\beta _A = 1.25$. 
The latent heat decreased by a factor of 
three as the lattice size was varied from $6^4$ to $16^4$. 
This information was extracted
from runs very close to the phase transition for each lattice, 
from which the 
latent heat was taken from the peak separation of the apparently 
bimodal distribution.
However, this method has a particular problem when used on the symmetric
lattice due to the symmetry actually being (Z2)$^4$ 
rather than just Z2 (assuming periodic boundary 
conditions in all four directions).
As the phase transition is approached on the finite lattice, the lattice 
goes from having four broken directions (and no unbroken), 
first to three broken
(and one unbroken), then to two broken, then one, and finally to the fully
unbroken symmetry case. The reason for this is simply the entropy factor
associated with each. There are 16 states with four broken directions,
$4\cdot 8=32 $ with three broken directions, 24 with two, and 8 with one and
only one corresponding state in the fully symmetric state. At the critical point
where these all have equal energy, the fully unbroken state would occur
only 1/81 of the time. It is easy to see how this state could be missed. If
one set $\beta$ so that this state occurred 50\% of the time, one would be below
the critical point (in $\beta$). The Boltzmann factor would then suppress the multiple-broken
direction cases, which might not appear at all on larger lattices where the energy
fluctuations are smaller. Thus, from a practical point of view, it is very difficult to
display the full range of symmetry-broken cases in a single simulation.
The multiple symmetry breakings appear 
to be associated with
nearly equal jumps in the plaquette (see Fig.~\ref{fig2}). 
On smaller lattices close to the transition, 
tunneling will occur between all of these states, showing nearly the full 
latent heat. However, larger lattices, with their smaller 
energy fluctuations
(requiring hitting the critical point more accurately) 
and longer tunneling times, may
only tunnel between two or three of the five levels in a reasonable-length
run, showing an apparently smaller
latent heat. 
Fig.~\ref{fig2}a shows a time history on an $8^4$ 
lattice at $\beta _A$=1.25
and $\beta _F$=1.2185 along with the Polyakov loop histories. 
There appear to be 
several energy plateaus in between the upper and lower, 
associated with only some of the Polyakov loop 
directions breaking. 
A similar simulation an a $12^4$ lattice with $\beta _F = 1.2183$ 
(closer to the 
critical point) populates only four of the five levels. 
Its plaquette histogram, shown in 
Fig.~\ref{fig2}b, shows four peaks. The missing peak in this case is 
from all-four 
Polyakov loops unbroken. 
A similar $16^4$ simulation at the same couplings populated 
only three peaks.
The multimodal nature of the distribution is not always as apparent 
as in Fig.~\ref{fig2}b.
If the peaks are unequally populated then only shoulders will be seen.
Thus it is risky to try to measure the latent heat from the widths of 
these distributions unless all five peaks of the multimodal distribution 
can be resolved. This multiple symmetry-breaking 
effect could easily explain the decrease in latent heat
with lattice size seen in Ref.~\cite{rajivsl}.

 The data presented in the current study are from 
hysteresis loops, which do not suffer from this 
problem, as not much time is spent
at $\beta _c$. As shown below, the hysteresis 
loops on $12^4$ and $20^4$ lattices are 
nearly identical, suggesting almost no finite size effects.  
Our values of latent
heat agree closely with the values Gavai and coworkers found on 
{\em asymmetric} lattices,
for which no finite size effect was found. Asymmetric 
lattices with one short
direction do not suffer
from the above problem, as the symmetry broken 
at the transition in question 
is then just 
the single Z2 of the short direction.

\section{Critical vs. Tricritical}
One needs to find an easily-measured quantity which can distinguish
the critical from the tricritical cases. 
As shown below, it turns out that the size of the hysteresis region, 
$(T^{**} - T^*)$ is
a linear function of the latent heat for the critical case
and a quadratic function for the tricritical case.
Here $T^*$ is the  lower (supercooling) metastability limit 
and $T^{**}$ is the upper (superheating). 
Between these temperatures, there are two minima 
of the free energy and tunneling exists. 
Outside of this region there is only one local minimum and
no tunneling exists.
This gross difference in scaling behavior follows from basic dimensional 
analysis of the powers in the Landau free energy,
but a detailed derivation is also given below. 
Both the latent heat and metastability regions are easily
determined from hysteresis sweeps. 
By plotting vs. latent heat no assumptions need be made concerning the 
possibly complex relationship between 
($\beta _F$, $\beta _A$) and the temperature and next higher coefficient 
in the free energy. It is interesting that this method is able to distinguish
between symmetry-breaking and non-symmetry-breaking first-order transitions
from energetics alone, without the need to identify the symmetry or order parameter.

In the Landau theory, the free energy is given 
by a power series in the order parameter.
\begin{equation}
f=\frac{1}{2} r \phi ^2 - w \phi ^3 + u_4 \phi ^4 + u_6 \phi ^6 .
\end{equation}
The quantity $r$ is an increasing  function of temperature, 
which can be defined as $r=a(T-T^* )$. 
For an ordinary 1st order transition that ends in a critical point, 
$w>0$ and $u_4>0$ ($w$ becomes zero at the 
critical point, whereas $u_4$ remains positive everywhere). 
The sixth order term can be ignored. 
The critical point occurs when $f=0$, $\partial f/ \partial \phi = 0$ for 
the minimum away from $\phi = 0$. 
These are easily solved for $\phi _c = w/2u_4$ and $r_c = w^2/2u_4$.
The latent heat can be obtained from the 
change in entropy between phases. Taking $s=-df/dT$ and expanding
$f$ to lowest order in $r$ about the two minima gives\cite{CL} 
$\Delta s=(a/2)\phi_{c}^2$. Therefore the latent heat is
given by 
\begin{equation}
q=T\Delta s = (aT_c /2)(w/2u_4)^2 .
\end{equation} 
Further details can be found in Ref.~\cite[pp.\ 168-175]{CL} from which
this and the following derivations are abstracted. 
Note that $q$ is quadratic in $w$, the parameter 
which is rapidly varying as one moves along the critical line, 
away from the critical point (rapidly varying 
because  it must vanish at the critical point). 
The metastability limit on superheating, $T^{**}$ occurs when the
local minimum away from $\phi =0$ 
becomes an inflection point instead, 
i.e. $\partial f/\partial \phi =0$ and
$\partial ^2 f /\partial \phi ^2 =0$. 
Solving these gives 
\begin{equation}
r^{**}=9w^2 /16u_4 ,
\end{equation}
also quadratic
in $w$. Taking the usual assumption that $u_4$ is slowly varying, 
results in the prediction that
\begin{equation}
T^{**}-T^*=a^{-1}r^{**} \propto q.
\end{equation}

In contrast, for the first-order transition that ends 
in a tricritical point, 
$w=0$ is enforced by symmetry. The quantity $u_4<0$ and 
one needs the positive $u_6$ term for stability. In this case {\em two} 
additional minima away from $\phi =0$ occur, one for 
positive and one for negative $\phi$. 
If these dip below the minimum at $\phi =0$ a first-order phase 
transition occurs. The tricritical point occurs when $u_4=0$. 
Beyond this is a line of second-order
transitions ($u_4 >0$). 
In this case the transition changes from first-order 
to second-order, rather 
than disappearing. In fact it must continue on, in order 
to divide the entire 
coupling plane into symmetry-broken and symmetry unbroken phases.  
Here $u_4$ is the rapidly
changing parameter as one moves along the 
transition line near the tricritical point, and
$u_6$ is assumed to be slowly varying. 
Following the same procedure given above
results in \cite{CL} 
\begin{equation}
\phi _c = \pm [ |u_4 |/2u_6 ]^{\frac{1}{2}}
\end{equation}
\begin{equation}
q=aT_c|u_4 |/(4u_6)
\end{equation}
(linear in $u_4$), 
and 
\begin{equation}
r^{**} =2u_4^2 /(3u_6)
\end{equation}
(quadratic in $u_4$).
Therefore the prediction for the 
tricritical case is 
\begin{equation}
T^{**}-T^* \propto q^2 .
\end{equation}

\section{Hysteresis Loops}

Rather slow hysteresis sweeps were performed to 
determine $q$ and $\Delta \beta \equiv \beta _F^{**} - \beta _F^*$,
for various fixed $\beta _A$. 
One has a fair degree of flexibility in deciding which 
parameter to choose as the 
temperature.  Here $\beta _F$ is being treated as the 
inverse (4-d) temperature in the partition function. 
The $\beta _A$ term with $\beta_ A$ held fixed can be thought 
of as either a temperature dependent external field
term (with coefficient $\beta_A / \beta _F$) or a temperature independent 
modification to the measure (contributing to the entropy). 
Because $\Delta \beta \ll \beta _F$, $\Delta \beta \propto T^{**} - T^*$ 
to lowest order.
Each run was begun with 500 equilibration sweeps, 
followed by 2000-4000 sweeps where $\beta _F$ is changed by 0.0001
on each sweep. This is slow enough that no 
hysteresis can be detected away from critical regions. 
Measurements were performed after each sweep. Runs were
performed on a $12^4$ lattice, 
except for five additional runs performed on a $20^4$ 
lattice to test for finite
size effects.  
Typical results for three different $\beta _A$ are 
shown in Fig.~\ref{fig3}. Each $12^4$ sweep was 
performed three times to test for repeatability and 
to estimate errors (only one is shown). 
The latent heat was measured as the 
jump in fundamental plaquette, $<\! p\!>$, at the 
position of maximum vertical
distance between
the hysteresis curves (with the definition of temperature above, 
the internal energy is given by $1-<\! p\!>$). 
The quantity $\Delta \beta$ was measured as the maximum width of the 
hysteresis curve.
One can also take $\beta _F ^*$ and $\beta _F ^{**}$ to be 
the points of maximum slope of the
hysteresis curves, and compute
$\Delta \beta$ from the difference - the results are nearly identical. 
Multiple runs are remarkably similar,
indicating modest statistical errors (detailed later). 
A worrisome systematic error from hysteresis sweeps is
the possibility of premature tunneling. 
If one sweeps too slowly, the system could tunnel to the other phase
before the metastability limit is reached. 
In rare instances it could happen a considerable distance away.
However, the fact that the $20^4$ sweeps were nearly identical 
to the $12^4$ at the same $\beta$'s 
(see Figs.~\ref{fig3} and \ref{fig4}), would seem to 
indicate this is not a problem. 
Tunneling times on the larger lattice are much longer, 
so if the smaller lattice
were tunneling prematurely by a significant amount 
then some difference between these different-size lattice runs
would be expected. The main result is shown in Fig.~\ref{fig4}, 
where $\Delta \beta$ is plotted against the square
of the jump in average plaquette, $(\Delta \! <\! p\!>)^2$ 
(proportional to $q^2$). 
Although three independent measurements for each point
are not sufficient to accurately determine individual error bars, 
an overall error estimate for the entire dataset
can be made, which indicates error bars of about one-third the
size of plotted points vertically and twice this horizontally.
If one does compute individual error bars, no particular trend is
observed - they are consistent with approximately equal error bars 
for all couplings.
A linear trend (which on these axes is a pure quadratic) 
is observed to fit the data well. 
Thus the data agree with the 
prediction of the tricritical case. 
If one tries to fit to a linear function of $\Delta \! <\! p\!>$, 
the result is
not satisfactory, with $\chi ^2/\rm{d.f.} =53$, 
whereas the pure quadratic shown in Fig.~\ref{fig4}, 
plotted as a straight line with axes given, 
has $\chi ^2 /\rm{d.f.}= 0.6$. A linear+quadratic fit was also performed.
In the possible case that the trend is linear, but the region of validity of
the Landau theory is small, this should be able to pick up the linear term.
This fit, however, gives a linear coefficient of $0.008 \pm 0.016$, consistent
with zero. The quadratic term (with coefficient $0.839\pm 0.039$) dominates
already at $(\Delta \! <\! p\!>)^2 = 0.0001$. Thus the data are not consistent with
``beginning linear" over any reasonable region of validity. 
Therefore, the data appear to be strongly inconsistent with the possibility
of there being an ordinary critical point at 
the FOE but entirely consistent with it being a tricritical point.

These results rely on certain assumptions 
inherent to the Landau theory, namely that higher order terms
in the free energy are slowly varying. 
However if this assumption were not valid it would be unlikely to obtain
such a clean result as pure quadratic scaling.  
Much more likely in this case would be a less conclusive result requiring a 
multiple-term fit.  Also, mean field theory is 
much more likely to give valid results in four dimensions than
in three, where it still provides useful predictions in many cases.

\section{Identification of the order parameter}

\begin{figure}[ht]
             \includegraphics[width=2.6in]{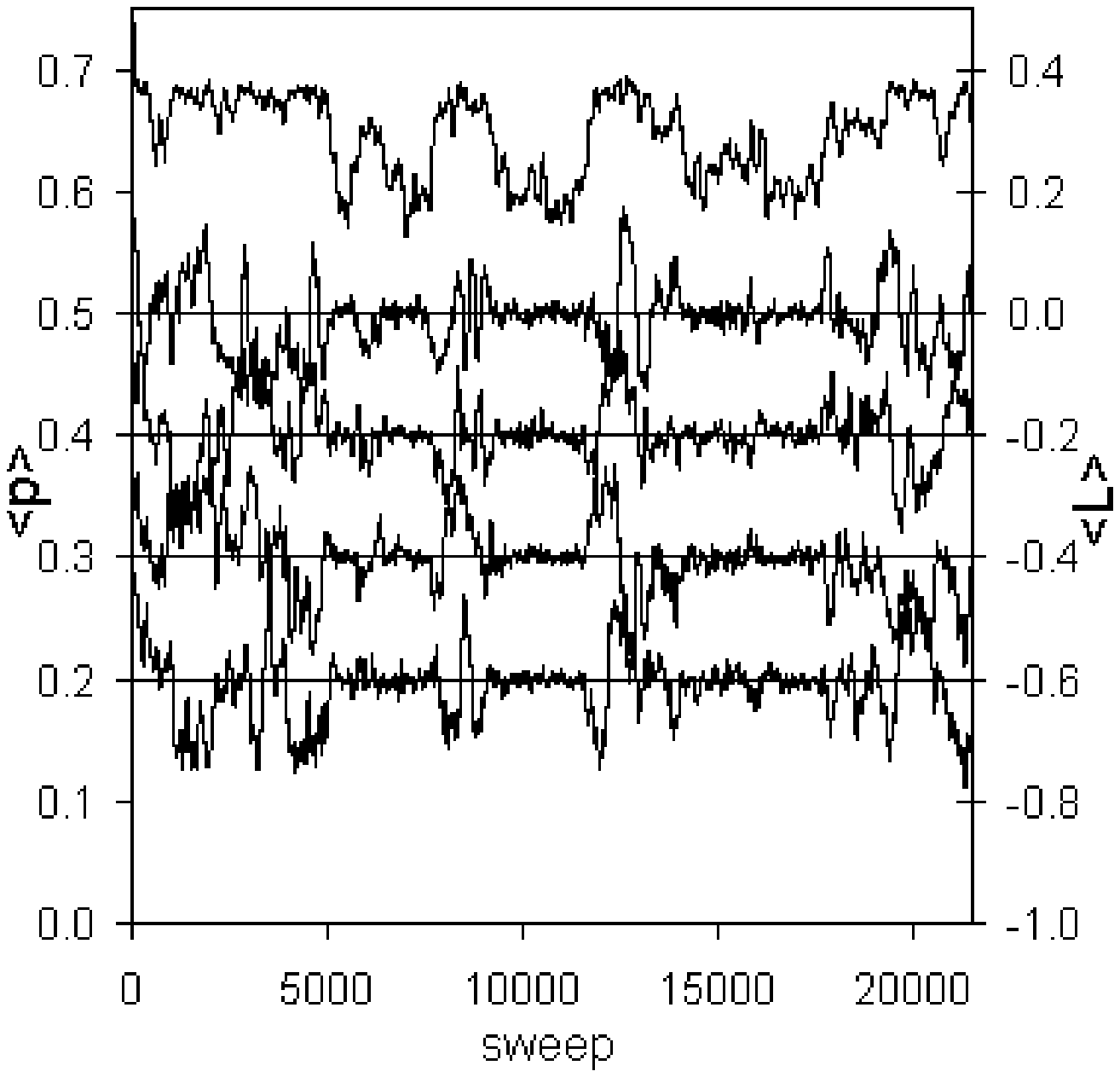} \quad
             \includegraphics[width=2.6in]{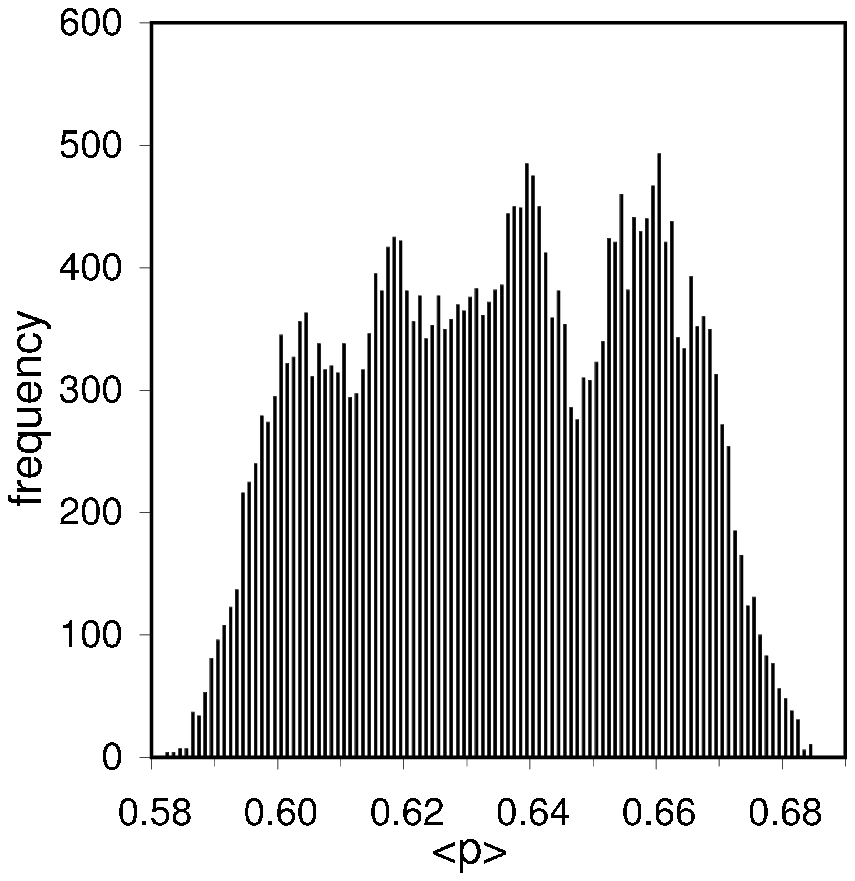} 
\caption{
(a) Time history of an $8^4$ lattice run at $\beta _F = 1.2185$, 
$\beta_A = 2.25$.
Upper trace is average plaquette, lower four traces are Ployakov loops.
Each successive Polyakov loop trace is offset by 0.2 for clarity. 
Steplike
structure in plaquette appears associated with the 
number of loops which show
spontaneous symmetry breaking at any time.
(b) Plaquette histogram on a $12^4$ lattice at $\beta _F = 1.2183$,
$\beta _A = 2.25$ showing clear multimodal distribution.}
\label{fig2}
\end{figure}

\begin{figure}[ht]
             \includegraphics[width=5in]{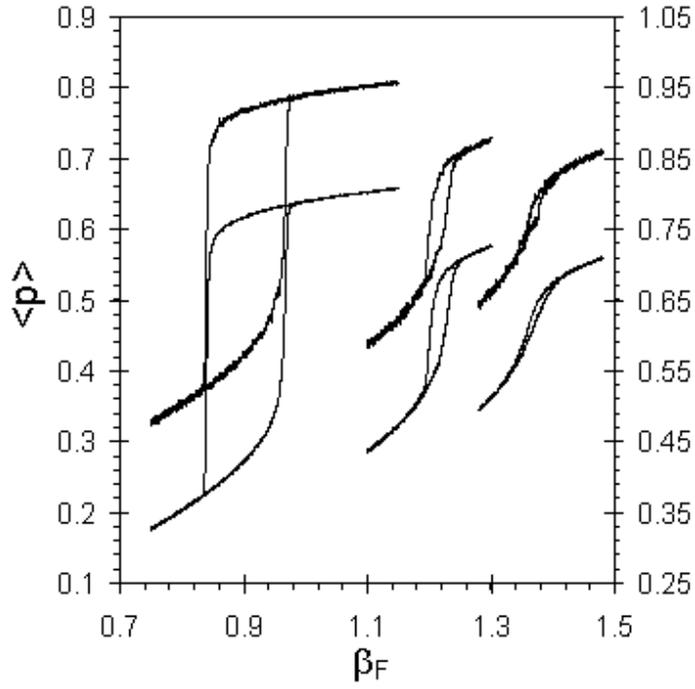}
\caption{
Hysteresis sweeps at $\beta _A$ = 1.7, 1.25, and 1.05. Upper curves
are $12^4$ lattice, lower offset curves are $20^4$ 
lattice (scale at right).
$\beta _F$ is changed by 0.0001 after each Monte-Carlo sweep. 
Near coincidence
of curves shows finite lattice size effects are small.} 
\label{fig3}
\end{figure}

\begin{figure}[ht]
             \includegraphics[width=5in]{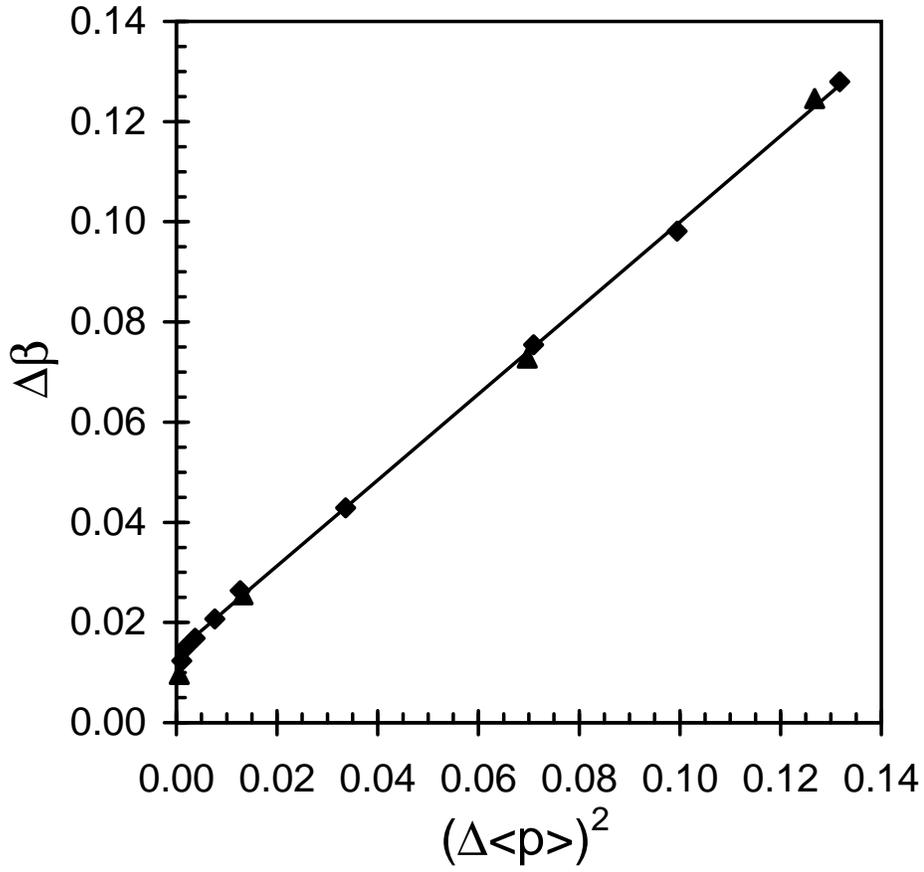}
\caption{
Width of metastable region vs. latent heat squared. Diamonds are
for $12^4$ lattice, triangles for $20^4$ lattice. Linear fit on these axes
is predicted by the
tricritical hypothesis. The alternative hypothesis, that of an 
ordinary critical point, predicts scaling directly with latent heat rather
than its square. The points comprise a 
range in $\beta _A$ from 1.05 to 1.7.}
\label{fig4}
\end{figure}

\begin{figure}[ht]
     \includegraphics[width=2.6in]{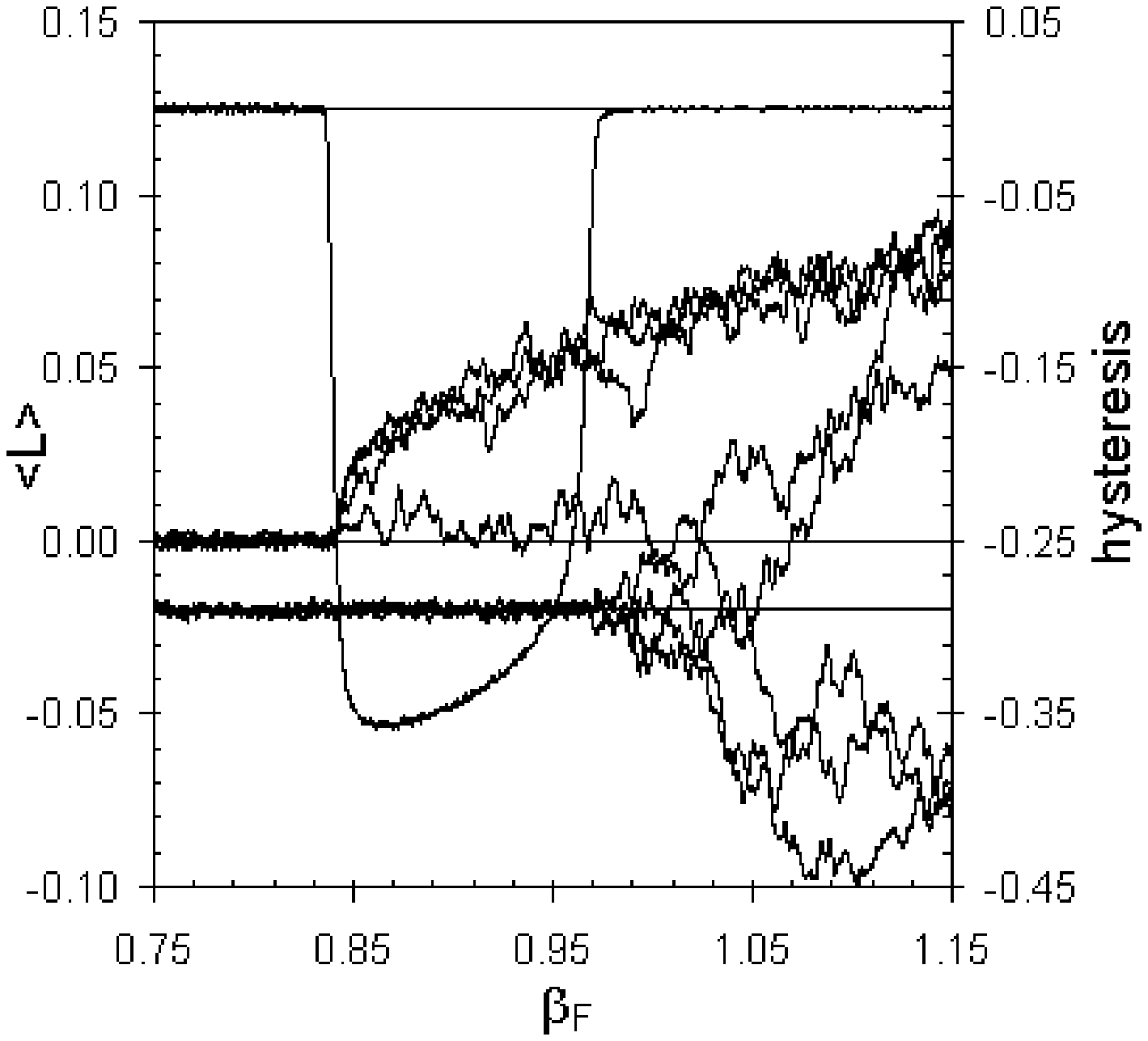}\hspace*{.5in}
\includegraphics[width=2.6in]{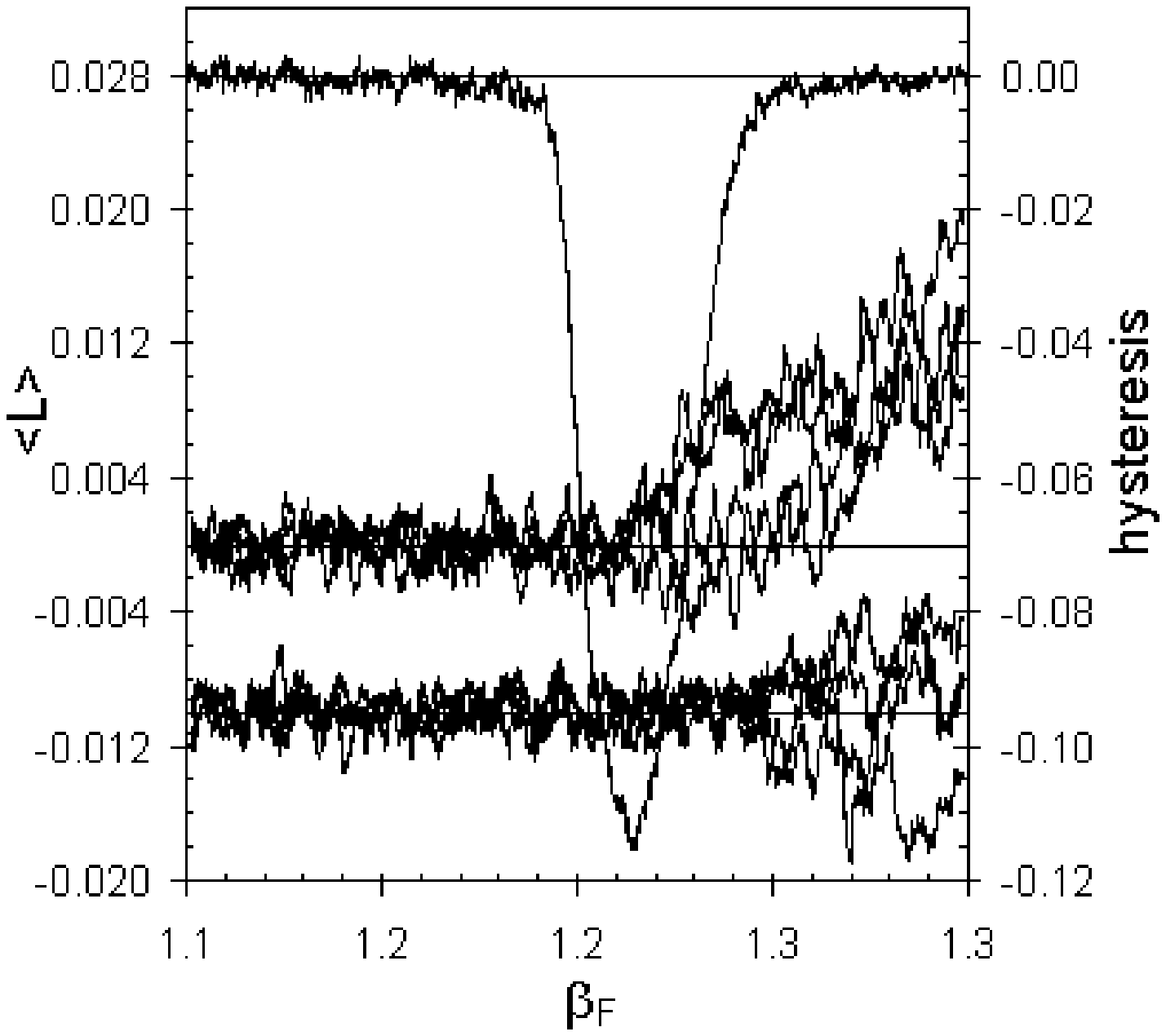}\vspace*{.4in}
\includegraphics[width=2.6in]{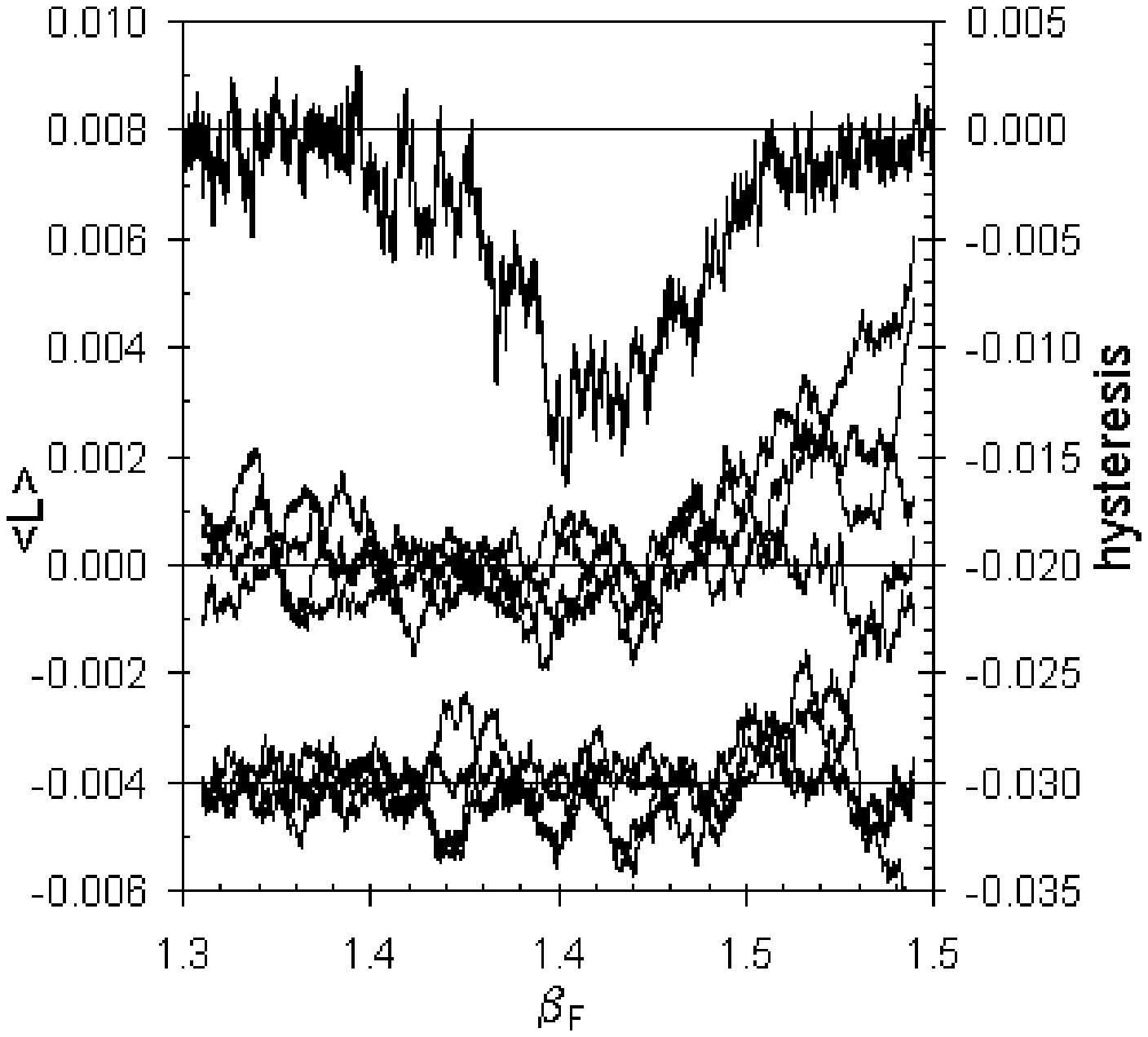}           
\caption{(a) A detailed look at $20^4$ hysteresis sweep at $\beta _A =1.7$.
Upper curve (right scale) is the difference in average 
plaquettes as measured on 
cooling vs. heating sweeps. Next four curves are Polyakov loops for heating
(beta decreasing) sweep. Final four curves, offset by 0.02 for clarity, are
Polyakov loops for cooling sweep. Polyakov loop curves are 25-point moving
averages to reduce variance. Spontaneous breaking of Polyakov loop is
clearly associated with plaquette tunneling events.
(b) Same at $\beta _A =1.25$.
(c) Same at $\beta _A = 1.0$. This lies below the tricritical point, in the 
second-order region. Now there is no longer much separation between 
the symmetry breakings for heating and cooling, but the 
breaking still seems to be
coincident with the rising edge of the plaquette hysteresis. Here 100-point
moving averages are used for the Polyakov loops. Note the hysteresis
curve has lost the steep sides associated with 
first-order tunneling, and its 
magnitude is small compared to the first-order values.}
\label{fig5}
\end{figure}

The tricritical behavior described above requires 
a symmetry breaking order parameter. So far, in this analysis, it has
not been necessary to identify this symmetry or the associated 
order parameter.  This was deliberately done in
order to make as few assumptions as possible. 
However, now that the tricritical case seems to be established
from energetics alone, it makes sense to try to identify 
the associated broken symmetry.  There are many reasons to
believe this is no other than the familiar 
Z2 Polyakov-loop symmetry. For one thing, this is the symmetry that
is broken in the attached Z2 lattice gauge theory at the top of 
the phase diagram. Secondly, the Polyakov loop
is seen to break along the bulk line, 
not just at the same couplings, 
but also tunneling at the same times. Fig.~\ref{fig5}
shows Polyakov-loop histories for heating and 
cooling sweeps together with plaquette 
histories for $\beta _A = 1.7$,
1.25 and 1.0 on the $20^4$ lattice.  These are well above, 
moderately above, and 
slightly below the tricritical point. In the first-order region, 
the metastability in Polyakov loop matches exactly
with that of the plaquette. However even in the 
second-order region the small hysteresis
signal in the plaquette from critical
slowing-down appears to be associated with the 
symmetry breaking of the Polyakov loop.
This, together with the correlations seen in Fig.~\ref{fig2} would seem to
indicate not merely coincident phase transitions, but an intimate 
locking of order parameters as well, since the tunnelings 
in plaquette and Polyakov loops are 
observed to take place at the same Monte Carlo times.

The $20^4$ data employ moving averages in the Polyakov loops to reduce the 
variance enough to see the symmetry breaking. The Polyakov loop
values in the broken region on such a large lattice are tiny. The 
individual datapoints 
are swamped by random fluctuations. Moving averages of, say 100 points,
reduce these random fluctuations by a factor of 10, allowing the small
nonzero average value in the broken phase to show through remarkably well.
Tunneling times in the broken region are generally much longer than this,
so the average values deduced are fairly accurate .

One important objection to the Polyakov loop being an order parameter
for a bulk transition is that its very definition depends on there
being periodic boundary conditions. A bulk transition, on the other hand,
should exist for any boundary conditions, such as open boundary conditions 
for which the Polyakov loop is not defined. However, it is important to 
realize that this same objection can be made for the Z2 lattice gauge
theory, for which there is no controversy about the existence of a 
bulk deconfining transition. Therefore, at least in this theory, there must be a 
second hidden 
order parameter and associated broken symmetry that exists for
the case of open boundary conditions. It is reasonable to expect that this
same situation may also exist in the SU(2) case. A possible symmetry
and order parameter which exist for both theories have been identified.
If one employs a partial axial gauge fixing that leaves unbroken a global
gauge symmetry on each 3-d layer of the lattice 
perpendicular to a certain direction,
then these remaining global gauge symmetries appear to break spontaneously at
weak coupling, apparently 
in concert with the Polyakov loop\cite{hidden}. The order
parameters are just the average perpendicular links in each layer.

The existence of a tricritical point implies that a 
bulk (4d zero-temperature) second-order transition exists below it,
down to and even below the Wilson axis (because it must divide the
plane into two non-connected regions with different symmetry). 
If the Polyakov loop is the order parameter then 
there is no finite-temperature
deconfinement transition. Deconfinement is a 
bulk transition and the zero-temperature continuum theory is not confining.
In order for the deconfinement transition to be a 
finite-temperature one as is usually assumed, it
must decouple from the bulk transition as described in the introduction. 
However, this has apparently not happened
yet, even on the $20^4$ lattice, in the coupling regions studied. 
If a separation does occur, then a {\em new} symmetry and
order parameter must be found for the second-order bulk transition
emanating from the tricritical point.  
It will be important to locate this 
new phase transition on the Wilson axis.
In order for lattice gauge results to be 
analytically connected to the continuum limit, 
one may only run simulations on the 
weak-coupling side of any bulk transition. 
If it occurs near the deconfinement transition on accessible lattices,
as is likely the case, then there would only be  
a narrow region of valid couplings in which reliable confining simulations
using the Wilson action could be run, 
lying between the new second-order bulk transition and the 
previously known finite-temperature transition. 
However, the 
fact that no second-order bulk transition separate 
from the deconfinement transition has ever been seen would
seem to make the entire scenario of two separate transitions unlikely. 
The order parameter studied in ref. \cite{hidden} breaks both
the global gauge symmetry and the Z2 Polyakov loop symmetry.
If this is driving the bulk transition, then it will only be
possible for the Polyakov loop deconfinement transition to
split off to the strong coupling side of this transition, because
on the weak coupling side the Z2 symmetry will already be broken by the bulk transition. 
In this case there would be no region of validity for the confining theory (always
separated from the continuum limit by the bulk transition). 

\begin{figure}[ht]
             \includegraphics[width=5in]{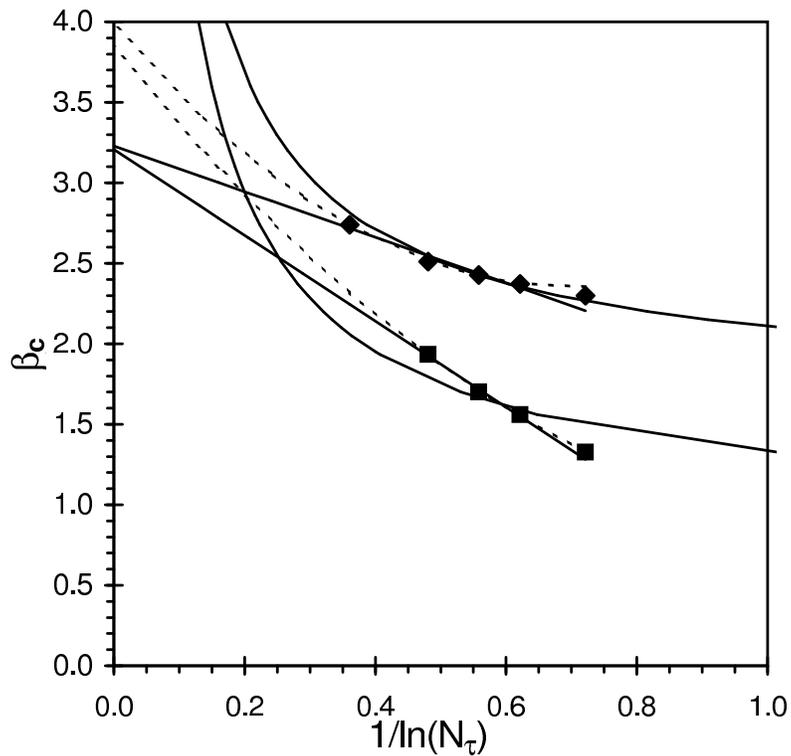}
\caption{Plot of $\beta _c$ for deconfinement transition on 
asymmetric lattices, to test
possibility of $1/\ln (N_{\tau})$ scaling law. 
Diamonds from Ref.~\cite{fhk}
are for Wilson action, squares from Ref.~\cite{rajivft} 
are for Z2 monopole and vortex suppressed action. 
Straight lines are linear fits to the data, excluding the $N_{\tau}=4$
points.
Dashed lines add a quadratic term to the fit.
Solid curved lines are, for comparison, 
fits to the 
normal two-loop scaling formula.}
\label{fig6}
\end{figure}

The evidence for the Polyakov-loop 
transition being a finite-temperature one stems 
mostly from an observed shift
in transition point with temporal lattice size, 
$N_\tau$, on asymmetric lattices with $N_ \sigma > N_ \tau$. The
size of the shift is larger than one usually expects 
for a bulk transition. However the possibility exists that
the four-dimensional non-abelian gauge theory could just 
have an unusually large finite-size shift. Fig.~\ref{fig6}
shows data for the finite-temperature 
deconfinement transition point, $\beta _c$ for the Wilson
action on different size asymmetric lattices
(data from Ref.~\cite{fhk}). Also shown is data from a 
different action used by Gavai\cite{rajivft}, in which Z2 monopoles
and vortices are suppressed. This action has the same 
$\Lambda$-parameter and therefore the same perturbative
scaling as the Wilson action, but the scaling is much 
different in the region of the deconfinement transition. 
Indeed, although the Wilson-action data appear to fit roughly to
the weak-coupling renormalization group scaling law (though
not acceptably, with a $\chi ^2 /d.f. = 28$), it does not fit at all to the 
Z2 monopole and vortex suppressed action data ($\chi ^2 /d.f. = 100$). However,
eliminating strong coupling lattice artifacts would be expected to 
improve scaling. This suggests that the rough fit of the Wilson-action data
may be accidental. A linear fit to $1/\ln (N_{\tau})$ also produces
a rough fit in the Wilson-action case (but also unacceptable considering  
the very small errors quoted for these points, with $\chi ^2 /d.f. = 64$).
A linear fit fares better with the Gavai data with $\chi ^2 /d.f. = 4.6$
(none of the fits discussed thus far include the lowest $N_{\tau}=4$ points).
The apparent intersection of these lines at $N_{\tau}=\infty$ is probably
fortuitous, but it is interesting that they would agree on the infinite lattice
critical point.
The scaling behavior of lattices with the same $\Lambda$-parameter
need only match in the weak coupling region. They do not need 
to match exactly at $\beta _{c \infty}$ (the critical coupling
for an infinite lattice), but since this $\beta$ is close
to the perturbative region they should be close.
Above a phase transition there 
is no reason for them to match at all, which could explain the 
rather different slopes. The dashed line fits include a quadratic
as well as a linear term (here the $N_{\tau}=4$ point was included in the Gavai-data
fit). This is able to accommodate the small curvature in the data rather well, and
still has near-agreement for the infinite lattice critical point of around $\beta = 3.9$.
The $\chi ^2 /d.f.$ for these fits are 3.5 and 2.3 which are coming rather close
to acceptability, considering the quadratic term is probably just approximating a
more complex non-linearity. A possible reason for approximate but not exact 
$1/\ln (N_{\tau})$
scaling is as follows.

The  behavior pictured in Fig.~\ref{fig6} can be 
understood if the transition is associated with  percolation of
abelian-monopole loops in the maximal abelian gauge. It has 
been shown that confinement seems to be associated with
the existence of a monopole loop that wraps through 
the periodic boundary, or with the closely-related existence
of a percolating cluster of such loops. The  deconfinement 
transition seems to occur when this is no longer the case.
The probability of a monopole loop of size $l$ (for $l<N$) existing on
a lattice, normalized per lattice site, has been shown to be 
proportional to $l^{-\gamma}$ where $\gamma$ is a $\beta$-dependent
quantity\cite{teper,myq}. Here $N$ is the linear size 
of a symmetric lattice. 
$\gamma$ is about 3 in the crossover region and becomes 
equal to 5 around $\beta = 2.9$\cite{myq}. The scaling law may
be somewhat different for $l>N$, but the results are insensitive to this
so long as the exponent $\gamma \ge 1$ for all $l>N$\cite{myq}.
To have a wrapping loop, one requires at least one loop of size
of order $N^{1+\epsilon}$ or larger, 
where $\epsilon$ is a fractal dimension
between 0 and 1, which has not yet been accurately measured. 
Probably $\epsilon >0$ because the loops are 
generally somewhat crumpled but $\epsilon = 0$ is 
also a possibility. The probability of finding such a loop
on an $N^4$ lattice is $CN^{(5+\epsilon )-\gamma (1+\epsilon)}$ 
where $C$ is 
some constant. This expression results from integrating the probability
over loop sizes beginning at the critical loop size for a wrapping loop.
Setting this probability to $\frac{1}{2}$ results in an 
estimate for the critical
value of $\gamma$ for 
that lattice, from which $\beta _c$ can be determined.
This therefore gives a model for the dependence of $\beta_c$ on $N$.
If one assumes that the asymmetric lattices, for which most of the 
data has been taken, have a similar scaling law, and also assuming
that $\gamma$ can be taken to be a linear function of $\beta$ in the 
region of interest (a simplification), then one obtains the 
finite-lattice scaling law 
\begin{equation}
\beta_c = \beta _{c\infty} -\frac{c}{ln(N)}
\end{equation} where $c$ is a constant.
This contrasts with the usual finite-lattice shift for a thermal 
transition 
\begin{equation}
\beta_c =\beta_{c \infty} -\frac{c}{N^{1/\nu}} 
\end{equation}
which converges much more rapidly to $\beta _{c \infty}$ as 
$N \rightarrow \infty$. 
The idea that this transition may be a 4-d percolation transition 
could explain why it has been so hard to identify, as there are not many
examples of such transitions. The fact that $\gamma$ is a somewhat
non-linear function of $\beta$ \cite{myq} could explain the need for a 
quadratic term in the fits. Better determination of $\gamma (\beta )$ and
measurements on asymmetric lattices would be needed
to come to a definitive conclusion on the higher-order terms.

It has been shown by Gavai and Mathur that the bulk 
first-order transitions are lattice artifacts that can be removed through
a judicious choice of action, 
leaving only a second-order deconfining transition\cite{gmla}. 
Indeed all bulk transitions are
lattice artifacts in that they do not affect the continuum limit.
However, the question still
exists whether this remaining second-order transition is 
bulk or finite-temperature. By using the original Bhanot-Creutz action,
one can learn more about this transition by its apparent 
connection to the first-order bulk transition at a 
tricritical point. The existence of a tricritical point,
surmised above from the behavior of the free energy around the
first-order transition, 
strongly implies that the second-order line is also bulk (i.e. a 4-d
zero-temperature transition), due to its attachment to
the bulk first-order endpoint. If this is true in the 
original action, then the continuum limit is deconfined for 
this action, and weak-coupling universality would imply 
this is also true for all other actions. Thus this work supports
the hypothesis made some time ago that the continuum limit
of SU(2) pure-glue lattice gauge theory may not 
be confining\cite{zp,ps2}.

One should, of course, remember that the SU(2) 
non-abelian gauge theory is not Quantum Chromodynamics (QCD) but an
approximation to it. Lack of confinement in the SU(2) theory 
does not imply that quarks are unconfined in the real world.
However, it does shed important light on 
the possible confinement mechanism. 
Of course this result must first 
be checked in the SU(3) case. If it holds up there, 
then suspicion must be cast on light quarks as the source 
of confinement in QCD, a position held by Gribov\cite{gribov,abgribov} and 
others\cite{zp,nc,cahill}. It could be that confinement is a byproduct of
chiral symmetry breaking rather than the other way around 
as sometimes stated. 
One possibility is that strong color fields disrupt the chiral condensate,
creating a bag of diminished and polarized chiral condensate 
around a hadron,
carrying an energy proportional to the volume of excluded condensate.
Supporting this conjecture is the observation in lattice simulations
that the strength of the chiral condensate is reduced from its vacuum value
in the presence
of a quark source\cite{markum}. 
 
A way to conclusively demonstrate confinement due to light
quarks would be to find an
action that completely erases the bulk confinement transition, 
including the second-order one, but agrees with the Wilson action
at weak coupling. Since all bulk transitions are lattice artifacts, 
this should be
possible. An action that
suppresses both Z2 monopoles and vortices as well as 
another gauge-invariant topological lattice artifact, 
the SO(3)-Z2 monopole \cite{so3z2}
appears promising. Simulations on lattices up to $30^4$
remain deconfined with this action for all couplings \cite{metoappear}. 
If confinement were to
return when light quarks are added to this theory, 
one would then have clear evidence of their essential role in 
confinement.

\newpage
\newpage
\noindent
\newpage
\newpage
\newpage
\newpage
\end{document}